\documentclass[aps,10pt,prl,twocolumn,showpacs,amsmath,amssymb]{revtex4}

\usepackage{graphicx}
\usepackage{dcolumn}   
\usepackage{bm}        
\usepackage{color}
\usepackage{flafter}  
\newcommand{\remove}[1]{{\color{blue}{\bf (removed)} #1}}
\renewcommand{\remove}[1]{}
\newcommand{\quiet}[1]{{\color{blue} #1}}
\renewcommand{\quiet}[1]{}
\newcommand{\comment}[1]{{\bf [Comment: #1 ]}}
\renewcommand{\comment}[1]{}
\newcommand{\green}{black}

\usepackage{amsmath}
\usepackage{amssymb}
\usepackage{makeidx}
\usepackage{graphicx}
\usepackage{color}

\newcommand{\hide}[1]{}

\newcommand{\myvec}[1]{\vec{#1}}

\newcommand{\vk}{{\myvec{k}}}

\newcommand{\val}{{\myvec{\alpha}}}

\newcommand{\vA}{{\myvec{A}}}

\newcommand{\cJ}{\mathcal{J}} %
\newcommand{\cP}{\mathcal{P}} 

\renewcommand{\r}{\rangle}
\renewcommand{\l}{\langle}

\newcommand{\om}{\omega}
\newcommand{\si}{\sigma}

\newcommand{\al}{\alpha}

\newcommand{\be}{\beta}
\newcommand{\Ga}{\Gamma}

\renewcommand{\bar}{\begin{array}{ll}}
\newcommand{\ear}{\end{array}}
\newcommand{\bma}{\begin{pmatrix}}
\newcommand{\ema}{\end{pmatrix}}
\newcommand{\beq}{\begin{equation}}
\newcommand{\eeq}{\end{equation}}
\newcommand{\bel}[1]{\begin{equation}\label{eq:#1}}
\newcommand{\eel}{\end{equation}}
\newcommand{\bea}{\begin{eqnarray}}
\newcommand{\eea}{\end{eqnarray}}

\newcounter{lecture}

\newcommand{\Ef}{\mathcal{E}}
\newcommand{\vEf}{\vec{\mathcal{E}}}

\begin{document}

\title{Anomalous Fano Profiles in External Fields}

\author{Alejandro Zielinski$^a$, Vinay Pramod Majety$^a$, 
Stefan Nagele$^b$, Renate Pazourek$^b$, Joachim Burgd\"orfer$^b$, and  Armin Scrinzi$^a$} \email{armin.scrinzi@lmu.de}
 \affiliation{$^a$Physics Department,
  Ludwig Maximilians Universit\"at, D-80333 Munich, Germany}
\affiliation{$^b$Institute for Theoretical Physics, Vienna Universtiy of Technology,Vienna, Austria}


\date{\today}

\begin{abstract}
We show that external control of Fano resonances in general leads to complex Fano $q$-parameters. 
Fano line shapes of photo-electron and transient absorption spectra in presence of an infrared control
field are investigated. Computed transient absorption spectra are compatible with a recent experiment 
[C. Ott {\it et al.}, Science 340, 716 (2013)] but suggest a modification of the interpretation proposed there. 
Control mechanisms for photo-electron spectra are exposed: control pulses applied {\em during} excitation modify 
the line shapes by momentum boosts of the continuum electrons.
Pulses arriving {\em after} excitation generate interference fringes due to infrared two-photon transitions. 
\end{abstract}

\maketitle
The celebrated Fano formula  \cite{Fano1961}
\beq\label{eq:fanoprofile}
\si(\Delta E)=\si_0\frac{|q\Gamma + 2\Delta E|^2}{\Gamma^2 + (2\Delta E)^2}
\eeq
describes the modulation of the cross section of any excitation process to a continuum that is structured
by a single embedded resonant state compared to the smooth background cross section $\si_0$ 
in absence of the embedded state. Apart from the resonance width $\Gamma$ and  the detuning $\Delta E$
there appears the  $q$ parameter, which produces a characteristic asymmetry
and --- if it is real --- an exact zero of the cross section. The Fano profile is one of the 
prominent manifestations of quantum mechanical interference in scattering. 
The mechanism is ubiquitous and independent of the particular nature of the transitions involved.
In recent years it was proposed to control the line shape by external fields and interactions, 
and schemes in diverse fields of physics were experimentally 
realized (see review in \cite{miroshnichenko10-fano-review}). For a quantum dot 
system controlled by a time-independent magnetic field it was observed that a generalization to complex 
$q$ was required to fit the control-dependence of the line shape.
Complex values of $q$ result from the breaking of time-reversal symmetry by the magnetic field \cite{kobayashi03-complexq}.
 In contrast, in standard Fano theory \cite{Fano1961},
applicable to time-reversal symmetric systems, $q$ is real-valued (see, e.g., \cite{lee99-time-reversal}). 
Complex $q$ has also been discussed as a signature of dephasing and
decoherence in atoms \cite{agarwal84-complexq,wickenhauser2005} as well as in quantum dots\cite{clerk01-fano-coherence}  
and microwave cavities \cite{baernthaler13:fano}.
More generally, complex $q$ are expected to appear whenever coupling to the environment or external fields turn the embedded state
into a state that cannot be described by a real-valued eigenfunction.

In this Letter we show that also a time-dependent electric control field, specifically an infrared (IR) probe pulse, 
generates complex $q$-parameters. Recently, the control of the line shape of transient absorption spectra (TAS) 
arising in a pump-probe scenario for helium was demonstrated \cite{Ott2013}:
the excitation of the {$2s\mathrm{n}p$} series of doubly excited states by a short extreme ultraviolet (XUV) pulse 
was probed by a weak, time-delayed near-IR pulse. The modulation of TAS line shapes was described as a control
of a real valued Fano $q$ parameter through an IR induced phase shift. Here we present  {\em ab initio} numerical solutions
that show that TAS as well as photo-electron spectra (PES) are characterized by complex rather than real $q$. 
For PES we  expose the two main mechanisms underlying the appearance of a non-zero  imaginary part of $q$ using a generalized Fano model
that includes an external control.  
First, we show that the phase shift discussed in \cite{Ott2013} directly leads to complex $q$ in PES. 
However, a second mechanism dominates the PES line shapes when XUV and IR 
pulses overlap: the free electron momenta are boosted by 
\beq\label{eq:streak}
\vk\to\vk+\vA(t_0),\quad\vA(t_0):=-\int_{t_0}^{t_1}\textcolor{\green}{\mathrm{d}\tau}\,\vEf(\tau),
\eeq
from their values after XUV excitation
time $t_0$ until the end of the IR field $\vEf(t)$ at time $t_1$. 
(Unless indicated otherwise, we use atomic units, where electron mass, 
proton charge, and $\hbar$ are all set equal to 1.)
The boost redistributes amplitudes among the partial waves and modifies the Fano interference
of the embedded state with the continuum in the $l=1$ decay channel.
Both mechanisms conserve the universal Fano line-shape Eq.~(\ref{eq:fanoprofile}), albeit with 
complex $q$.

An important higher order process that leads to a departure from the Fano profile is 
two-IR-photon coupling, which was discussed for a multiplet of embedded states \cite{zhao2005-fano} 
and for Autler-Townes splitting \cite{Chu2013}. In the present setting, 
two-IR-photon coupling generates characteristic interference-like modulations of the PES
when the XUV and IR pulses are well-separated in time. Similar structures will appear
when the decaying state is partially depleted by a delayed IR \cite{zhao12_fano}. The importance of 
non-resonant IR multi-photon processes in the excitation of dipole-forbidden auto-ionizing states 
was noted recently \cite{chu14-dipole-forbidden}. 

\begin{figure}
  \includegraphics[width=1.\linewidth]{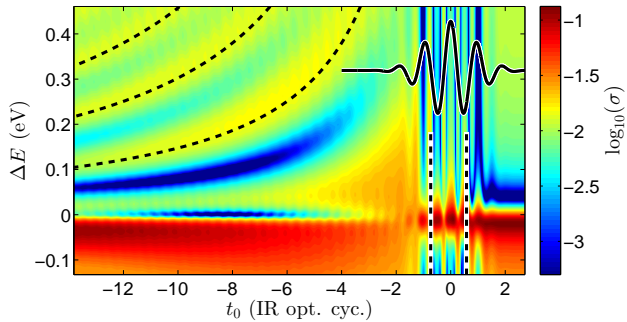}
\caption{\label{fig:overview}PES $\si(\Delta E)$ in the vicinity of the 2s2p resonance 
as a function \textcolor{\green}{of} XUV excitation time $t_0$ ($l=1$ partial wave). 
IR peak intensity $2\times10^{12}W/cm^2$. Solid line: IR field,
curved fine-dashed lines: $2n\pi\hbar/|t_0|, n=1,2,3$, closely follow interference maxima.
Negative $t_0$ correpsond to the XUV preceding IR.
Dashed vertical lines indicate the lineout times of Fig.~\ref{fig:lineout}.
}
\end{figure}

We first present the PES in the vicinity of the He(2s2p) line as a 
function of XUV excitation time $t_0$ with an IR pulse centered at $t=0$ (Fig.~\ref{fig:overview}). 
The results were obtained by numerically solving the 
time-dependent Schr\"odinger equation of the He atom in full 3+3 spatial dimensions.
Spectra were computed using the time-dependent surface flux method (tSURFF, \cite{tao12:ecs-spectra,Scrinzi2012}). 
For a 
summary of the computational approach and discussion of its accuracy, see \cite{majety15:hacc}.
The XUV center  wavelength of $\lambda=21\,nm$ was chosen to match the excitation to the $2s2p$ state, but the spectral width of 
$\sim10\,eV$ at the pulse duration of
\textcolor{\green}{$0.15\,fs$} 
 evenly covers the entire $2s\mathrm{n}p$ series of doubly excited He states.
The calculations were performed for IR wavelength of $800\,nm$ with a pulse duration of 2 optical cycles,
peak intensity $2\times10^{12}W/cm^2$, and parallel linear polarization of the XUV and IR pulses. 

\begin{figure}
  \includegraphics[width=0.85\linewidth]{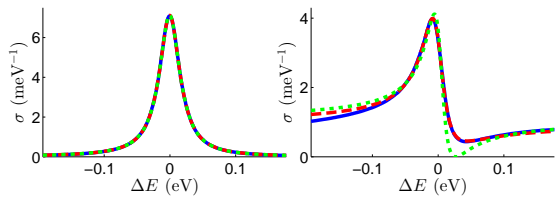}
\caption{\label{fig:lineout}
PES at two different delay times as indicated in Fig.~\ref{fig:overview}.
\remove{Resonance position is at $\Delta E=0$.} 
Left: $t_0=1/2$ IR opt.cyc., near a field peak, right: $t_0=-3/4$, near a field node. IR peak intensity $10^{12}W/cm^2$.
Solid: numerical result, dashed: fit admitting complex $q$, dot-dashed: fit with $q$ restricted to real. 
All three lines nearly coincide at $t_0=1/2$. 
\comment{ finer lines, make colors consistent with Fig.5, choose black for numerical result}
}
\end{figure}

The two lineouts (Fig.~\ref{fig:lineout}) pertain to $t_0=-3/4$ 
and $+1/2$ (in units of IR optical cycle). At $t_0=-3/4$, where the XUV 
pulse coincides with a node of the IR electric field, one sees a strongly asymmetric Fano profile. In contrast, at 
$t=+1/2$, peak of the IR field, the profile is Lorentz-like. 
Neither the width nor the position of the resonance are affected by the weak IR. The pattern is repeated as 
$t_0$ is scanned through the IR pulse. 

Fig.~\ref{fig:lineout} also contains fits to the lines by Eq.~(\ref{eq:fanoprofile}), where $\Gamma$ was taken 
from the IR-free case. Only the overall intensity $\si_0$ and the $q$-parameter and were adjusted, either 
restricting $q$ to real values and 
or admitting complex $q$, respectively. When the XUV coincides with a node of the IR (right panel of Fig.~\ref{fig:lineout}),
only the fit with complex $q$ is satisfactory: there is no exact zero in the spectrum when 
IR and XUV pulse overlap, which trivially rules out an accurate fit by Eq.~(\ref{eq:fanoprofile}) with real $q$. 

For the example of PES we show how complex $q$ arises in the framework of a generalized Fano theory.
A standard Fano Hamiltonian has the form
\beq\label{eq:fanomodel}
H= |\varphi\r E_\varphi \l\varphi|+\!\!\int|\vk\r\frac{k^2}{2}\l \vk |+ |\vk\r V_\vk\l\varphi|+ |\varphi\r V_\vk^*\l\vk| \mathrm{d}^3 k,
\eeq
where the embedded bound state $|\varphi\r$ interacts with the continuum states $|\vk\r$
through $V_\vk=\l \vk | V |\varphi\r$.
Solutions are known for the exact scattering eigenfunctions $|\xi_\vk\r$, the resonance width $\Ga$, and the shift 
of the resonance position from the non-interacting $E_\varphi$. 
The Fano transition amplitude  $\l \xi_\vk | T| \phi_0\r$ 
for an arbitrary transition operator $T$ from some initial
state $|\phi_0\r$ leads to the Fano cross section (\ref{eq:fanoprofile}).
We introduce the wave packet after transition
\beq
T|\phi_0\r=:|\psi_0\r=|\varphi\r X_\varphi+ \int \textcolor{\green}{\mathrm{d}}^3k \,|\vk\r X_\vk,
\eeq
with the transition amplitudes from the initial state
$X_\varphi=\l\varphi|T|\phi_0\r$ and $X_\vk=\l\vk|T|\phi_0\r$. 
For notational simplicity we consider the case where $|\varphi\r$ decays into a well-defined angular momentum state,
in case of the 2s2p doubly excited state this is the $l=1$ partial wave. 
The $q$-parameter for the standard Fano Hamiltonian (\ref{eq:fanomodel}), denoted as $q_0$, is
\beq
q_0=\frac{1}{\pi V_k^* k}
\frac{\l\varphi|\psi_0\r + \cP\int k'^2 \textcolor{\green}{\mathrm{d}}k'\frac{2V^*_{k'}}{k^2-k'^2}\l k'|\psi_0\r}
{\l k | \psi_0\r},
\eeq
where the $|k\r$ denotes the $l=1$ partial wave continuum states with $k=\sqrt{\vk^2}$ and $V_k$ are the  
coupling matrix elements between embedded and continuum states. 
When the $\l k'|\psi_0\r$ and $\l \varphi |\psi_0\r$ all share the same phase, $q$ is real. 
\hide{This is the case 
in the standard Fano model \cite{Fano1961}, where all participating states $|\phi_0\r$, $|\varphi\r$ and
$|k\r$ are assumed to be stationary and can be chosen to be real. In addition, the transition $T$ is taken to 
be ``sudden'', which implies that the transition matrix elements do not carry an energy-dependent phase.
}

In our model for the transition in the pulse overlap region, we assume that the initial state $|\phi_0\r$ is unaffected
by the IR and that the effect on the embedded 
state $|\varphi\r$ is only a Stark shift $\Delta E_\varphi(t)$ relative to the field-free energy $E_\varphi$. 
The interaction of the IR with the continuum states is described in 
the standard ``strong field approximation'' \cite{lewenstein94:harmonics}: when the IR field prevails over the atomic potential,
the continuum states at time $t$ can be approximated as plane waves with wave vector $\vk-\vA(t)$
and the phase of the time-evolution is modified accordingly.
Finally, we assume that the
IR pulse duration is short compared to the decay time of the embedded state. With that the net effect
of the IR pulse is to replace $|\psi_0\r$ by a modified initial wave packet 
\beq\label{eq:psi1}
|\psi_1\r=|\varphi\r X_\varphi+\int \textcolor{\green}{\mathrm{d}}^3k\,
\textcolor{\green}{\mathrm{e}}^{-\textcolor{\green}{\mathrm{i}}\Phi_\vk}|\vk\r X_{\vk-\vA(t_0)}
\eeq
(see Supplemental Material \cite{supp:fano}).
The phase offset $\Phi_\vk$ between embedded and continuum states accumulated 
from excitation at $t_0$ until the end of the IR pulse at $t_1$, $\vA(t_1)=0$, is
\beq
\textcolor{\green}{\Phi_{\vk}={ \int_{t_{0}}^{t_{1}}\mathrm{d}\tau}\,\Big([\vk-\vA(\tau)]^{2}/2-E_\varphi -\Delta E_{\varphi}(\tau)\Big)}.
\eeq
Clearly, even if the initial amplitudes $X_\varphi$ and $X_{\vk-\vA(t_0)}$ are all real, the interaction with the IR imprints
a phase-modulation on $|\psi_1\r$ and the Fano parameter becomes complex. Moreover, in
$X_{\vk-\vA(t_0)}$ the partial waves are redistributed compared to the IR-free $X_\vk$
by the addition of a streaking momentum $\vA(t_0)$. A short
calculation (see \cite{supp:fano}) 
leads to the IR modification of the Fano parameter
\beq\label{eq:q1}
q_1=q_0+a \left(\mathrm{e}^{-\mathrm{i}\chi}/\cJ-1\right),
\eeq
where $a=\l \varphi |\psi_0\r/(\pi V^*_kk\l k|\psi_0\r)$ denotes the ratio
of embedded to continuum amplitudes without the IR, and
\beq\label{eq:chi}
\textcolor{\green}{\chi=\int_{t_{0}}^{t_{1}}\mathrm{d}\tau\,\left[\Delta E_{\varphi}(\tau)-\vec{A}^{2}(\tau)/2\right]}
\eeq
 is a laser-induced phase shift between the two components. 
Although the phase-shift $\chi$ does give a numerically discernable contribution,
the $t_0$-dependence of $q_1$ is dominated by 
\beq\label{eq:Jfactor}
\cJ=j_0(|\val|k)-2j_2(|\val|k)-3\mathrm{i}\frac{j_1\left(|\val|k\right)}{|\val|k}\,\val\cdot\vA(t_0), 
\eeq
where the spatial offset of a free electron by the IR pulse 
\beq\label{eq:spatialoffset}
\val=\int_{t_0}^{t_1} \textcolor{\green}{\mathrm{d}\tau}\,\vA(\tau)
\eeq 
appears in the argument of the spherical Bessel functions $j_l$.
The $\cJ$-term accounts for streaking by 
the term $\int_{t_0}^{t_1} \vk\cdot\vA(\tau)d\tau$ in $\Phi_\vk$.
By the dipole selection rule $|\psi_0\r$ has angular momentum $l=1$ and therefore only the $j_0$, $j_1$, and $j_2$ contribute
to the $l=1$ partial wave emission. 

\begin{figure}
\includegraphics[width=0.85\linewidth]{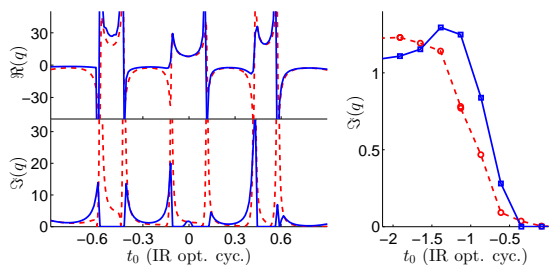}
\caption{\label{fig:complexq} Dependence of $q_1$ on XUV-IR delay time $t_0$ according to 
Eq.~(\ref{eq:q1}) (dashed lines) and from fits to numerical results (solid lines). 
Upper panel: $\Re(q_1)$, lower panel: $\Im(q_1)$.
Right: $\Im(q_1)$ at times $t_n$ near the nodes of the field ($\val=0$, see text): 
Eq.~(\ref{eq:q1}) (bullets) and fit (squares).
}
\end{figure}

In Fig.~\ref{fig:complexq} we compare Eq.~(\ref{eq:q1}) with fits to the numerically computed 2s2p line.
For the fits, the amplitude ratio was kept constant at the field-free value of $a=-3.3$ and Stark shifts 
were neglected, $\Delta E_\varphi\equiv0$. 
Sign changes of the real part and peaks in the imaginary parts
are all well reproduced. Quantitative deviations must be expected in the strong field approximation, 
for example, due to the use of plane waves $|\vec{k}\r$ instead of the exact scattering solutions.
In addition, there is a non-negligible
IR two-photon coupling, as will be discussed below.

There are excitation times $t_0=t_n$ where the spatial offset
vanishes, $\val=0$, and therefore $\cJ=1$. At these delays, the imaginary part of $q_1$ is exclusively due to the phase-shifts $\chi$. 
Up to small corrections arising from the short IR pulse duration, the $t_n$ coincide with 
zeros of the field. At the $t_n$ the profile is Fano-like (Fig.~\ref{fig:lineout}),
except that the characteristic minimum remains slightly above zero.  
In our model, the minima for subsequent $t_n$'s grow monotonically as the delay $|t_0|$ increases 
(Fig.~\ref{fig:complexq}, right panel) reflecting the accumulation of the shift $\chi$, Eq.~(\ref{eq:chi}).
For overlapping pulses, all resonances $2s\mathrm{n}p$, $n=2$ through 7 show the same delay-dependence, 
which corroborates that line-shape modulations are dominated by the dressing of the continuum.

When the XUV precedes the IR pulse without overlap (large negative $t_0$), 
the spatial offset goes to zero ($\val\approx0$) and therefore
$\cJ\approx 1$. 
Here, line-shapes are the combined effect of the phase-shift $\chi$ and IR two-photon coupling between 
embedded and continuum states. Two-photon coupling is not included in the standard Fano model Eq.~(\ref{eq:fanomodel}). 
It manifests itself in side-band like interferences. 
Stimulated 2-IR-photon emission creates ripples in the otherwise smooth non-resonant PES around the 
energies $E_{-}=E_\varphi - 2\om$, where $\om$ is the IR photon energy. 
The electron amplitude generated by two-photon absorption near $ E_{+}= E_\varphi+2\om$ is super-imposed 
with the higher-lying Fano resonances and therefore not clearly discernable.   
Absorption-emission transitions couple the embedded state to the continuum near $ E=E_\varphi$. 
We model the spectral features near  $E_\be,\,\be=\varphi$ and $\be=-$ by  
\bea\label{eq:interference}
\si( E)&=&|f(E) + \Ef_0^2 e^{-it_0 (E-E_\be)}b_\be(E)|^2,
\eea
where $t_0$ is the XUV-IR delay, $\Ef_0$ denotes the IR peak field strength, and
$f(E)$ is the spectral amplitude in absence of the IR. The unknown two-IR-photon transition amplitudes
are parameterized as   $b_\be(E)=c_\be g(E-E_\be)$. For $g$ we use a Gaussian profile with a fixed 
width equal to the spectral width of the IR. The only adjustable parameters are the two-photon 
coupling strengths $c_\be$, accounting for the different strengths of the transition into structured
and unstructured continuum.

\begin{figure}
  \includegraphics[width=0.85\linewidth]{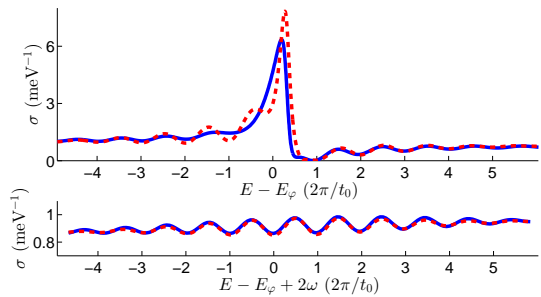}
\caption{\label{fig:wiggles} Two-photon interference resonance in the $l=1$ partial wave near 
$E_\varphi$ (upper panel) and near $E_\varphi-2\om$ (lower). 
Cross-section Eq.~(\ref{eq:interference}) (dot-dashed) is compared to the full numerical result (solid) at $t_0=-12$ opt.cyc.
}
\end{figure}

In Fig.~\ref{fig:wiggles} the cross-section Eq.~(\ref{eq:interference}) at
$t_0=-12$\comment{check sign} IR opt.cyc.\ is compared to the TDSE result. 
The fringe separation of $2\pi/t_0$ discernable in  Figs.\ref{fig:overview} and \ref{fig:wiggles} proves that the structures
are caused  by interference of
photo-electrons emitted at relative delay $t_0$. Without any further adjustment of $c_\be$ or $g$ the model
equally well reproduces the spectra for varying intensities up to $I\lesssim10^{12}W/cm^2$ and for all $t_0$.
The quadratic dependence on IR field strength $\Ef_0$ shows that this is a true two-photon process 
without resonant coupling to neighboring states.
At short time-delays $t_0$ the effect is negligible, as fringe separation diverges and 
fringes are hardly discernable. 
\remove{Note that the effect
is ``heterodyning'', i.e.\ a re-distribution of probability generated by the original 
Fano resonance, little extra electron probability added by the coupling.}

\begin{figure}
\includegraphics[width=0.95\linewidth]{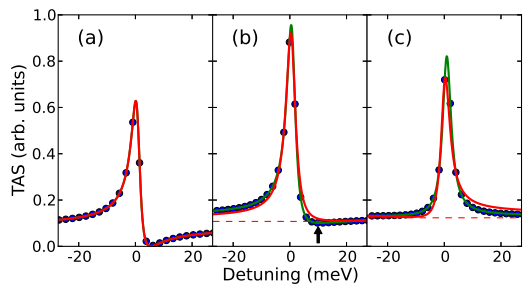}
\caption{\label{fig:tas} TAS at the $2s4p$ resonance. 
Results are shown for (a) XUV only, (b) a \textcolor{\green}{$7\,fs$} FWHM IR pulse with 
peak intensity $I=2\times10^{12}W/cm^2$ reached at arrival of the XUV pulse, and (c) the peak IR intensity
\textcolor{\green}{$5\,fs$} after the XUV. Dots: numerical results; green lines: fit of a Fano profile with complex $q$
(arrow marks the local minimum), 
solid red lines: Fano profile with real $q$ according \cite{Ott2013} and additional offset indicated by the dashed
red line.   
}
\end{figure}

Turning now to the complementary channels for observing Fano line shapes by the IR field we calculate
TAS  near the \textcolor{\green}{$2s\mathrm{n}p$}, $n\leq6$ states
using the pulse parameters of Ref.~\cite{Ott2013}.
The spectra were determined from the full TDSE solutions following the method described in Ref.~\cite{gaarde11:absorption}. 

The delay dependence of the TAS differs from that of the 
PES: under the influence of the IR field the characteristic minimum of the
Fano line all but disappears and does not fully reappear for 
delays shorter than the resonance life time. Fig.~\ref{fig:tas} shows as an example the 
computed transient absorption line of the $n=4$ resonance at different IR delays.
In all cases the numerical results conform with a generalized Fano line with a complex $q$ 
parameter. For comparison, we include the Fano shapes with the purely real $q$-parameter predicted 
in \cite{Ott2013}, where a constant pedestal of absorption was admitted to account for the offset 
of the lines from zero. The point to be noted is that the present numerical calculation is background-free, 
i.e., such a pedestal cannot be explained by a background contribution from other channels.
 The weak local minimum seen in the numerically computed absorption line at 
delay 0 corresponds to complex $q=-1.2+i2.4$. This is at variance with real valued $q=-9$ resulting from the model
proposed in \cite{Ott2013}.  
Both, the offset from zero and the absorption minimum in the vicinity of the
peak indicate that the influence of the control field is not fully captured by the model of \cite{Ott2013}.

The complex $q$ in TAS and PES appears at parameters that are accessible 
by experimental setups as in Ref.~\cite{Ott2013}. For either observable
the signature of complex $q$ is a local minimum above 0. For PES
the shape should follow Eq.~(\ref{eq:q1}), for TAS the exact shape can be obtained numerically.
The main experimental difficulty obviously is the proper background
subtraction. In case of the PES, the presence of the IR 
introduces a smooth background of partial waves, Eq.~(\ref{eq:streak}), in addition to the 
$l=1$ partial wave that exhibits the Fano interference. However,
at the times $t_n$ where $\vec{\al}=0$ the contributions
from the other partial waves are negligible and $\Im(q)$ as given in the right panel in 
Fig.~\ref{fig:complexq} is directly observable in the angle-integrated cross section. At other delays,
the $l=1$ cross section must be reconstructed from an angle-resolved measurement (see, e.g., \cite{garcia04:inversion}).

In summary, anomalous Fano profiles with complex $q$-parameter appear 
whenever a non-trivial relative phase between embedded state and continuum 
is imprinted on the system during the Fano decay. Such a phase can reflect internal dynamics 
of the embedded state $|\varphi\r$, i.e.\ when it is not strictly an eigenstate of a stationary Hamiltonian, 
as for decaying states and de-coherence. It can equally be generated by an external control, as demonstrated
here. Our theoretical description of the process should be generalizable to systems where 
we can model the impact of the control on bound and embedded states and when
control time is short compared to the resonance life time. This is the case for laser 
pulses on atoms or molecules, but the approach is also valid, e.g., for time-dependent 
electric or magnetic fields acting on quantum dots.  

We acknowledge support by the excellence cluster ``Munich
Center for Advanced Photonics (MAP)'' and by the Austrian Science
Foundation project ViCoM (F41) and NEXTLITE (F049).

\bibliography{fano.bib}

\begin{thebibliography}{20}
\expandafter\ifx\csname natexlab\endcsname\relax\def\natexlab#1{#1}\fi
\expandafter\ifx\csname bibnamefont\endcsname\relax
  \def\bibnamefont#1{#1}\fi
\expandafter\ifx\csname bibfnamefont\endcsname\relax
  \def\bibfnamefont#1{#1}\fi
\expandafter\ifx\csname citenamefont\endcsname\relax
  \def\citenamefont#1{#1}\fi
\expandafter\ifx\csname url\endcsname\relax
  \def\url#1{\texttt{#1}}\fi
\expandafter\ifx\csname urlprefix\endcsname\relax\def\urlprefix{URL }\fi
\providecommand{\bibinfo}[2]{#2}
\providecommand{\eprint}[2][]{\url{#2}}

\bibitem[{\citenamefont{Fano}(1961)}]{Fano1961}
\bibinfo{author}{\bibfnamefont{U.}~\bibnamefont{Fano}}, \bibinfo{journal}{Phys.
  Rev.} \textbf{\bibinfo{volume}{124}}, \bibinfo{pages}{1866}
  (\bibinfo{year}{1961}).

\bibitem[{\citenamefont{Miroshnichenko
  et~al.}(2010)\citenamefont{Miroshnichenko, Flach, and
  Kivshar}}]{miroshnichenko10-fano-review}
\bibinfo{author}{\bibfnamefont{A.~E.} \bibnamefont{Miroshnichenko}},
  \bibinfo{author}{\bibfnamefont{S.}~\bibnamefont{Flach}}, \bibnamefont{and}
  \bibinfo{author}{\bibfnamefont{Y.~S.} \bibnamefont{Kivshar}},
  \bibinfo{journal}{Rev. Mod. Phys.} \textbf{\bibinfo{volume}{82}},
  \bibinfo{pages}{2257} (\bibinfo{year}{2010}),
  \urlprefix\url{http://link.aps.org/doi/10.1103/RevModPhys.82.2257}.

\bibitem[{\citenamefont{Kobayashi et~al.}(2003)\citenamefont{Kobayashi, Aikawa,
  Katsumoto, and Iye}}]{kobayashi03-complexq}
\bibinfo{author}{\bibfnamefont{K.}~\bibnamefont{Kobayashi}},
  \bibinfo{author}{\bibfnamefont{H.}~\bibnamefont{Aikawa}},
  \bibinfo{author}{\bibfnamefont{S.}~\bibnamefont{Katsumoto}},
  \bibnamefont{and} \bibinfo{author}{\bibfnamefont{Y.}~\bibnamefont{Iye}},
  \bibinfo{journal}{Phys. Rev. B} \textbf{\bibinfo{volume}{68}},
  \bibinfo{pages}{235304} (\bibinfo{year}{2003}),
  \urlprefix\url{http://link.aps.org/doi/10.1103/PhysRevB.68.235304}.

\bibitem[{\citenamefont{Lee}(1999)}]{lee99-time-reversal}
\bibinfo{author}{\bibfnamefont{H.-W.} \bibnamefont{Lee}},
  \bibinfo{journal}{Phys. Rev. Lett.} \textbf{\bibinfo{volume}{82}},
  \bibinfo{pages}{2358} (\bibinfo{year}{1999}),
  \urlprefix\url{http://link.aps.org/doi/10.1103/PhysRevLett.82.2358}.

\bibitem[{\citenamefont{Agarwal et~al.}(1984)\citenamefont{Agarwal, Haan, and
  Cooper}}]{agarwal84-complexq}
\bibinfo{author}{\bibfnamefont{G.~S.} \bibnamefont{Agarwal}},
  \bibinfo{author}{\bibfnamefont{S.~L.} \bibnamefont{Haan}}, \bibnamefont{and}
  \bibinfo{author}{\bibfnamefont{J.}~\bibnamefont{Cooper}},
  \bibinfo{journal}{Phys. Rev. A} \textbf{\bibinfo{volume}{29}},
  \bibinfo{pages}{2552} (\bibinfo{year}{1984}),
  \urlprefix\url{http://link.aps.org/doi/10.1103/PhysRevA.29.2552}.

\bibitem[{\citenamefont{Wickenhauser et~al.}(2005)\citenamefont{Wickenhauser,
  Burgdoerfer, Krausz, and Drescher}}]{wickenhauser2005}
\bibinfo{author}{\bibfnamefont{M.}~\bibnamefont{Wickenhauser}},
  \bibinfo{author}{\bibfnamefont{J.}~\bibnamefont{Burgdoerfer}},
  \bibinfo{author}{\bibfnamefont{F.}~\bibnamefont{Krausz}}, \bibnamefont{and}
  \bibinfo{author}{\bibfnamefont{M.}~\bibnamefont{Drescher}},
  \bibinfo{journal}{Phys. Rev. Lett.} \textbf{\bibinfo{volume}{94}},
  \bibinfo{pages}{023002} (\bibinfo{year}{2005}).

\bibitem[{\citenamefont{Clerk et~al.}(2001)\citenamefont{Clerk, Waintal, and
  Brouwer}}]{clerk01-fano-coherence}
\bibinfo{author}{\bibfnamefont{A.~A.} \bibnamefont{Clerk}},
  \bibinfo{author}{\bibfnamefont{X.}~\bibnamefont{Waintal}}, \bibnamefont{and}
  \bibinfo{author}{\bibfnamefont{P.~W.} \bibnamefont{Brouwer}},
  \bibinfo{journal}{Phys. Rev. Lett.} \textbf{\bibinfo{volume}{86}},
  \bibinfo{pages}{4636} (\bibinfo{year}{2001}),
  \urlprefix\url{http://link.aps.org/doi/10.1103/PhysRevLett.86.4636}.

\bibitem[{\citenamefont{B\"arnthaler et~al.}(2010)\citenamefont{B\"arnthaler,
  Rotter, Libisch, Burgd\"orfer, Gehler, Kuhl, and
  St\"ockmann}}]{baernthaler13:fano}
\bibinfo{author}{\bibfnamefont{A.}~\bibnamefont{B\"arnthaler}},
  \bibinfo{author}{\bibfnamefont{S.}~\bibnamefont{Rotter}},
  \bibinfo{author}{\bibfnamefont{F.}~\bibnamefont{Libisch}},
  \bibinfo{author}{\bibfnamefont{J.}~\bibnamefont{Burgd\"orfer}},
  \bibinfo{author}{\bibfnamefont{S.}~\bibnamefont{Gehler}},
  \bibinfo{author}{\bibfnamefont{U.}~\bibnamefont{Kuhl}}, \bibnamefont{and}
  \bibinfo{author}{\bibfnamefont{H.-J.} \bibnamefont{St\"ockmann}},
  \bibinfo{journal}{Phys. Rev. Lett.} \textbf{\bibinfo{volume}{105}},
  \bibinfo{pages}{056801} (\bibinfo{year}{2010}),
  \urlprefix\url{http://link.aps.org/doi/10.1103/PhysRevLett.105.056801}.

\bibitem[{\citenamefont{Ott et~al.}(2013)\citenamefont{Ott, Kaldun, Raith,
  Meyer, Laux, Evers, Keitel, Greene, and Pfeifer}}]{Ott2013}
\bibinfo{author}{\bibfnamefont{C.}~\bibnamefont{Ott}},
  \bibinfo{author}{\bibfnamefont{A.}~\bibnamefont{Kaldun}},
  \bibinfo{author}{\bibfnamefont{P.}~\bibnamefont{Raith}},
  \bibinfo{author}{\bibfnamefont{K.}~\bibnamefont{Meyer}},
  \bibinfo{author}{\bibfnamefont{M.}~\bibnamefont{Laux}},
  \bibinfo{author}{\bibfnamefont{J.}~\bibnamefont{Evers}},
  \bibinfo{author}{\bibfnamefont{C.~H.} \bibnamefont{Keitel}},
  \bibinfo{author}{\bibfnamefont{C.~H.} \bibnamefont{Greene}},
  \bibnamefont{and} \bibinfo{author}{\bibfnamefont{T.}~\bibnamefont{Pfeifer}},
  \bibinfo{journal}{Science} \textbf{\bibinfo{volume}{340}},
  \bibinfo{pages}{716} (\bibinfo{year}{2013}).

\bibitem[{\citenamefont{Zhao and Lin}(2005)}]{zhao2005-fano}
\bibinfo{author}{\bibfnamefont{Z.~X.} \bibnamefont{Zhao}} \bibnamefont{and}
  \bibinfo{author}{\bibfnamefont{C.~D.} \bibnamefont{Lin}},
  \bibinfo{journal}{Phys. Rev. A} \textbf{\bibinfo{volume}{71}},
  \bibinfo{pages}{060702} (\bibinfo{year}{2005}),
  \urlprefix\url{http://link.aps.org/doi/10.1103/PhysRevA.71.060702}.

\bibitem[{\citenamefont{Chu and Lin}(2013)}]{Chu2013}
\bibinfo{author}{\bibfnamefont{W.-C.} \bibnamefont{Chu}} \bibnamefont{and}
  \bibinfo{author}{\bibfnamefont{C.~D.} \bibnamefont{Lin}},
  \bibinfo{journal}{Phys. Rev. A} \textbf{\bibinfo{volume}{87}},
  \bibinfo{pages}{013415} (\bibinfo{year}{2013}).

\bibitem[{\citenamefont{Zhao and Lein}(2012)}]{zhao12_fano}
\bibinfo{author}{\bibfnamefont{J.}~\bibnamefont{Zhao}} \bibnamefont{and}
  \bibinfo{author}{\bibfnamefont{M.}~\bibnamefont{Lein}}, \bibinfo{journal}{New
  Journal of Physics} \textbf{\bibinfo{volume}{14}}, \bibinfo{pages}{065003}
  (\bibinfo{year}{2012}).

\bibitem[{\citenamefont{Chu et~al.}(2014)\citenamefont{Chu, Morishita, and
  Lin}}]{chu14-dipole-forbidden}
\bibinfo{author}{\bibfnamefont{W.-C.} \bibnamefont{Chu}},
  \bibinfo{author}{\bibfnamefont{T.}~\bibnamefont{Morishita}},
  \bibnamefont{and} \bibinfo{author}{\bibfnamefont{C.~D.} \bibnamefont{Lin}},
  \bibinfo{journal}{Phys. Rev. A} \textbf{\bibinfo{volume}{89}},
  \bibinfo{pages}{033427} (\bibinfo{year}{2014}),
  \urlprefix\url{http://link.aps.org/doi/10.1103/PhysRevA.89.033427}.

\bibitem[{\citenamefont{Tao and Scrinzi}(2012)}]{tao12:ecs-spectra}
\bibinfo{author}{\bibfnamefont{L.}~\bibnamefont{Tao}} \bibnamefont{and}
  \bibinfo{author}{\bibfnamefont{A.}~\bibnamefont{Scrinzi}},
  \bibinfo{journal}{New Journal of Physics} \textbf{\bibinfo{volume}{14}},
  \bibinfo{pages}{013021} (\bibinfo{year}{2012}).

\bibitem[{\citenamefont{Scrinzi}(2012)}]{Scrinzi2012}
\bibinfo{author}{\bibfnamefont{A.}~\bibnamefont{Scrinzi}},
  \bibinfo{journal}{New Journal of Physics} \textbf{\bibinfo{volume}{14}},
  \bibinfo{pages}{085008} (\bibinfo{year}{2012}), ISSN
  \bibinfo{issn}{1367-2630}.

\bibitem[{\citenamefont{Majety et~al.}(2015)\citenamefont{Majety, Zielinski,
  and Scrinzi}}]{majety15:hacc}
\bibinfo{author}{\bibfnamefont{V.~P.} \bibnamefont{Majety}},
  \bibinfo{author}{\bibfnamefont{A.}~\bibnamefont{Zielinski}},
  \bibnamefont{and} \bibinfo{author}{\bibfnamefont{A.}~\bibnamefont{Scrinzi}},
  \bibinfo{journal}{New Journal of Physics} \textbf{\bibinfo{volume}{17}},
  \bibinfo{pages}{063002} (\bibinfo{year}{2015}),
  \urlprefix\url{http://stacks.iop.org/1367-2630/17/i=6/a=063002}.

\bibitem[{\citenamefont{Lewenstein et~al.}(1994)\citenamefont{Lewenstein,
  Balcou, Ivanov, L'Huillier, and Corkum}}]{lewenstein94:harmonics}
\bibinfo{author}{\bibfnamefont{M.}~\bibnamefont{Lewenstein}},
  \bibinfo{author}{\bibfnamefont{P.}~\bibnamefont{Balcou}},
  \bibinfo{author}{\bibfnamefont{M.~Y.} \bibnamefont{Ivanov}},
  \bibinfo{author}{\bibfnamefont{A.}~\bibnamefont{L'Huillier}},
  \bibnamefont{and} \bibinfo{author}{\bibfnamefont{P.~B.}
  \bibnamefont{Corkum}}, \bibinfo{journal}{Phys. Rev. A}
  \textbf{\bibinfo{volume}{49}}, \bibinfo{pages}{2117} (\bibinfo{year}{1994}).

\bibitem[{sup()}]{supp:fano}
\emph{\bibinfo{title}{Supplemental material: derivation of eq.~(8).}}

\bibitem[{\citenamefont{Gaarde et~al.}(2011)\citenamefont{Gaarde, Buth, Tate,
  and Schafer}}]{gaarde11:absorption}
\bibinfo{author}{\bibfnamefont{M.~B.} \bibnamefont{Gaarde}},
  \bibinfo{author}{\bibfnamefont{C.}~\bibnamefont{Buth}},
  \bibinfo{author}{\bibfnamefont{J.~L.} \bibnamefont{Tate}}, \bibnamefont{and}
  \bibinfo{author}{\bibfnamefont{K.~J.} \bibnamefont{Schafer}},
  \bibinfo{journal}{Phys. Rev. A} \textbf{\bibinfo{volume}{83}},
  \bibinfo{pages}{013419} (\bibinfo{year}{2011}),
  \urlprefix\url{http://link.aps.org/doi/10.1103/PhysRevA.83.013419}.

\bibitem[{\citenamefont{Garcia et~al.}(2004)\citenamefont{Garcia, Nahon, and
  Powis}}]{garcia04:inversion}
\bibinfo{author}{\bibfnamefont{G.~A.} \bibnamefont{Garcia}},
  \bibinfo{author}{\bibfnamefont{L.}~\bibnamefont{Nahon}}, \bibnamefont{and}
  \bibinfo{author}{\bibfnamefont{I.}~\bibnamefont{Powis}},
  \bibinfo{journal}{Rev. Sci. Instrum.} \textbf{\bibinfo{volume}{75}},
  \bibinfo{pages}{4989} (\bibinfo{year}{2004}).

\end{thebibliography}

\end{document}


\title{Supplemental Material: derivation of equation (8)}

\author{Alejandro Zielinski$^a$, Vinay Pramod Majety$^a$, 
Stefan Nagele$^b$, Renate Pazourek$^b$, Joachim Burgd\"orfer$^b$, and  Armin Scrinzi$^a$}
 \affiliation{$^a$Physics Department,
  Ludwig Maximilians Universit\"at, D-80333 Munich, Germany}
\affiliation{$^b$Technische Universit\"at Wien, A-1040 Vienna, Austria}


\date{\today}

\maketitle

For describing the action of the IR on the continuum we use the ``strong field approximation'', 
where the field dominates the dynamics and continuum states are assumed to be unaffected by the scattering potential.
Only the phase modification and momentum boost by the external dipole field is 
taken into account. The laser pulse is described through $\vA(t)$, Eq.~(2) of the main 
text, and $t_1$ denotes the end of the pulse where $\vA(t_1)=0$.
In this approximation, the time-evolution in the IR dipole field from 
a state with momentum $\vk-\vA(t_0)$ at $t_0$ until the end of the pulse at $t_1$, where it has the final momentum $\vk$, is given as
\begin{equation}
U_{\text{IR}}|\vk-\vA(t_0)\r= e^{-i\int_{t_0}^{t_1} dt  [\vk-\vA(t)]^2/2}|\vk\r. 
\end{equation}
The effect of the IR on the embedded state is assumed to be a modification of its phase according to the
Stark-shifted energy $E_\varphi(t)=E_\varphi+\Delta E_\varphi(t)$:
\beq
|\varphi,t_1\r=\mathrm{e}^{-\mathrm{i}\int_{t_{0}}^{t_{1}}\mathrm{d}t\,\left[E_\varphi+\Delta E_{\varphi}(t)\right]}|\varphi,t_0\r.
\eeq
If we neglect the coupling $V_\vk$ between $|\varphi\r$ and $|\vk\r$ for the duration of the IR, we can insert
these time-evolutions into Eq.~(4) of the main text to obtain the modification of the initial wave packet $|\psi_0\r$ by the IR pulse
in the time interval from its creation by the XUV at $t_0$ until $t_1$
\begin{equation}\label{eq:psi1t}
|\tilde{\psi}_1\rangle=
|\varphi\r\mathrm{e}^{-\mathrm{i}\int_{t_{0}}^{t_{1}}\mathrm{d}t\, \left[E_{\varphi}+\Delta E_\varphi(t)\right]}X_{\varphi}
+\int\mathrm{d}^{3}k\,\mathrm{e}^{-\mathrm{i}\int_{t_{0}}^{t_{1}}\mathrm{d}t\,\left[\vec{k}-\vA(t)\right]^{2}/2}
|\vk\r X_{\vk-\vA(t_0)}.
\end{equation}
 $|\tilde{\psi}_1\r$ equals the initial wavepacket $|\psi_1\r$ of Eq.~(6) of the main text up to the overall phase
$\exp\left\{-\mathrm{i}\int_{t_{0}}^{t_{1}}\mathrm{d}t\left[E_\varphi+ \Delta E_{\varphi}(t)\right]\right\} $.
As any global phase does not affect the crossection, we will drop the distinction in the following steps.
The dipole transition matrix element from an initial $l=0$ state 
has the general form $X_\vk=(\hat{\epsilon}\cdot\vk) Z_{|\vk|}$ 
for polarization direction $\hat{\epsilon}$.
We neglect the $|\vk|$-dependence of $Z_{|\vk|}\equiv Z$, 
as the resonance width $\Gamma$ is very narrow compared to spectral width of $\sim 12\,eV$
of an $0.15\,fs$ XUV pulse,
and also the energy dependence of the transition to the structureless continuum states $|\vk\r$ far from 
threshold can be considered negligible across $\Gamma$. 

We introduce the IR-induced offset $\val:=\int_{t_0}^{t_1}\vA(t) dt$,
define the polarization direction as the $z$-axis,
and write the expansion of a plane wave into spherical Bessel functions $j_l$ and Legendre polynomials $P_l$ as
\beq
\mathrm{e}^{\mathrm{i}k\al\cos\th_k}=\sum_{l}\mathrm{i}^{l}(2l+1)P_l(\cos\th_k)j_{l}\big(k\al\big),
\eeq
where $\th_k$ is the angle of $\vk$ with the $z$-axis.

We denote by $|Y_1^0\r$ the spherical harmonics for the spatial coordinates and assume $l=1$ symmetry
for $|\varphi\r=|Y_1^0\r|\varphi_1\r$.
For the $l=1$ partial wave of $|\psi_1\r$  one can perform the integration over angles of $\vk$
\begin{eqnarray}
|Y_1^0\r\l Y_1^0|\psi_1\rangle & = & e^{-i(t_1-t_0)E_\varphi}|\varphi\r
X_\varphi+\int dk\,k^2 e^{-i(t_1-t_0)k^2/2}|k\r\sqrt{\frac{4\pi}{3}}kZ e^{-i\chi}\cJ(k),
\end{eqnarray}
where $|k\r=|Y_l^0\r\l Y_1^0|\vk\r$ is the $l=1$ partial wave of the continuum state and 
$\chi$ and  $\cJ$ are defined by Eqs.~(9) and (10) of the main text.

\comment{
Details of the calculation (to be removed in the final version):
\beq
|\vk\r = \sum_{l'm}Y^{m*}_l(\uk)|Y_l^m\r|k\r_l
\eeq
\beq
X_{\vk-\va}=k\cos\th_k - A
\eeq
\beq
\xi:=\int_{t_{0}}^{t_{1}}\mathrm{d}t\,k^2/2+A^2/2
\eeq
Close with $\l Y_1^0|$ (only $\vk$ part)
\beq
\int dk\,k^2 \int d\cos\th \int d\varphi \,e^{-i\xi}
\sum_{l}i^l(2l+1)P_l(\cos\th_k)j_{l}(k\al)
\sqrt{\frac{3}{4\pi}}\cos\th_k|k\r_1 (k\cos\th_k - \hat{\epsilon}\cdot\vA)Z
\eeq
For checking the factors, $j_0(k\al)$ term:
\beq
=\int dk\,k^2 \sqrt{\frac{4\pi}{3}}  \,e^{-i\xi}
j_0(k\al)|k\r_1 k Z
\eeq
$j_1(k\al)$ term:
\bea
&&
\int dk\,k^2 \int d\cos\th \int d\varphi \,e^{-i\xi}
i(3)\cos\th_kj_{1}(k\al)
\sqrt{\frac{3}{4\pi}}\cos\th_k|k\r_1 ( - \hat{\epsilon}\cdot{\vA})Z
\\
&=&\int dk\,k^2 \frac{4\pi}{1} \,e^{-i\xi}
ij_{1}(k\al)
\sqrt{\frac{3}{4\pi}}|k\r_1 ( - \hat{\epsilon}\cdot\vA)Z
\eea

}

The $q$-parameter for the modified wave-packet $|\psi_1\r$ replacing the initial wave packet $|\psi_0\r$ is 
\begin{equation}
q_{1}=\frac{1}{\pi V_{k}^{*}k}\frac{\langle\Phi|\psi_1\rangle}{\langle k|\psi_1\rangle}.
\end{equation}
with 
\beq
|\Phi\r:=|\varphi\r + \cP\int k'^2 \mathrm{d}k'|k'\r\frac{2V^*_{k'}}{k^2-k'^2},
\eeq
compare Eq.~(5) of the main text.
In the evaluation of the principal value integral
we neglect the $k$-dependence of $\cJ$  
\begin{equation}
\mathcal{P}\int k'^{2}\mathrm{d}k'\,\frac{V_{k'}^{*}}{E_{k}-E_{k'}}
\l k'|\psi_{1}\r
\approx\mathcal{J}(k)\mathrm{e}^{-\frac{\mathrm{i}}{2}\int_{t_0}^{t_1} dt\,A^{2}(t)}\mathcal{P}\int k'^{2}\mathrm{d}k'\, 
\frac{V_{k'}^{*}}{E_{k}-E_{k'}}e^{-i(t_1-t_0)k^2/2}\l k'|\psi_{0}\r.
\end{equation}
Note that $\l k'|\psi_{0}\r=\frac{4\pi}{3}k'Z$.
We find for the $l=1$ component
\begin{eqnarray}
q_{1} & = & \frac{1}{\pi V_{k}^{*}k}\left\{\frac{\l\Phi|U_0|\psi_{0}\r}{\l k|U_0|\psi_{0}\r}
+\frac{\langle\varphi|U_0|\psi_{0}\rangle}{\langle k|U_0|\psi_{0}\rangle}
\left[\mathrm{e}^{-\mathrm{i}\chi}/\mathcal{J}(k)-1\right]\right\}.
\end{eqnarray}
$U_0$ is the free time-evolution during the time interval $[t_0,t_1]$ in absence of the IR and without the coupling $V_k\equiv0$.
It is given by $U_0|k\r=e^{-i(t_0-t_1)k^2/2}|k\r$ and  $U_0|\varphi\r=e^{-i(t_1-t_0) E_\varphi}|\varphi\r$.
The effect of $U_0$ on the $q$-parameter is negligible for time-intervals that are short
compared to the decay time: $t_1-t_0\ll 1/\Gamma$:
\beq\label{eq:neglectV}
\frac{\langle\Phi|\psi_{0}\rangle}{\langle k|\psi_{0}\rangle}
\approx \frac{\langle\Phi|U_{0}|\psi_{0}\rangle}{\langle k|U_{0}|\psi_{0}\rangle}\quad\text{ and }\quad
\frac{\langle\varphi|\psi_{0}\rangle}{\langle k|\psi_{0}\rangle}
\approx \frac{\langle\varphi|U_{0}|\psi_{0}\rangle}{\langle k|U_{0}|\psi_{0}\rangle}.
\eeq
With this the first term can be identified with $q_{0}$ and we have obtained Eq.~(8) of the 
main article.